\documentclass[%
notitlepage,
amsmath,amssymb,
10pt,
]{revtex4-1}

\usepackage{graphicx}
\usepackage{dcolumn}
\usepackage{bm}
\usepackage{bbold}
\usepackage{braket}
\usepackage{mathtools}
\usepackage{setspace}
\usepackage[colorinlistoftodos]{todonotes}

\begin{document}

\preprint{APS/123-QED}
\title{Point-coupling Hamiltonian for frequency-independent linear optical devices}

\author{Rahul Trivedi$^\xi$}
\author{Kevin Fischer$^\xi$}
\author{Sattwik Deb Mishra}
\author{Jelena Vu\v{c}kovi\'c}
\affiliation{%
E. L. Ginzton Laboratory, Stanford University, Stanford, CA 94305, USA  \\
$^\xi$ Both the authors contributed equally to this work}%
\date{\today}

\begin{abstract}
We present the point-coupling Hamiltonian as a model for frequency-independent linear optical devices acting on propagating optical modes described as a continua of harmonic oscillators. We formally integrate the Heisenberg equations of motion for this Hamiltonian, calculate its quantum scattering matrix, and show that an application of the quantum scattering matrix on an input state is equivalent to applying the inverse of classical scattering matrix on the annihilation operators describing the optical modes. We show how to construct the point-coupling Hamiltonian corresponding to a general linear optical device described by a classical scattering matrix, and provide examples of Hamiltonians for some commonly used linear optical devices. Finally, in order to demonstrate the practical utility of the point-coupling Hamiltonian, we use it to rigorously formulate a matrix-product-state based simulation for time-delayed feedback systems wherein the feedback is provided by a linear optical device described by a scattering matrix as opposed to a hard boundary condition (e.g.~a mirror with less than unity reflectivity).
\end{abstract}

\maketitle

\section{Introduction}
\noindent Linear-optical elements are key components in any quantum information processing system \cite{blais2007quantum, imamog1999quantum, duan2001long}. They can be used for applying phase shifts, interfering, and splitting quantum light propagating in waveguide modes or collimated optical beams. Consequently, they are traditionally described by scattering matrices \cite{fan2003temporal, ruan2009temporal}. The response of the linear optical element to an incoming state is often calculated by applying the inverse of the classical scattering matrix on the annihilation operators \cite{mandel1995optical, kok2007linear, skaar2004quantum}. {The full time evolution of a quantum state incident on linear-optical elements has also been captured within the framework of quantum stochastic differential equations \cite{gough2009series, petersen2010quantum, parthasarathy2012introduction}. Several attempts have been made to calculate a Hamiltonian that captures the dynamics of linear optical elements \cite{prasad1987quantum, fearn1987quantum,garcia2016multiple, leonhardt2003explicit}. In particular, Refs.~\cite{leonhardt2003explicit} and \cite{garcia2016multiple} show the existence and explicit construction of an anti-hermitian matrix whose exponential reproduces the classical scattering matrix of a linear-optical device. A Hamiltonian describing the quantum physics of the linear-optical device can then be constructed from this matrix. However, this approach needs to introduce a fictitious time corresponding to the duration for which the incident photons interact with the linear-optical device. While the introduction of this fictitious time does not impact the relationship between the input and output states, it makes the model unphysical if the full dynamics of the quantum state is desired. Alternatively, as shown in Ref.~\cite{dalton1999quasi}, a physically accurate Hamiltonian for the linear-optical device can be written down if the normal modes of the linear-optical device are known.} \\

\noindent In this paper, we propose a Hamiltonian that can model an arbitrary frequency-independent linear optical device acting on propagating optical modes. The Hamiltonian assumes that the optical modes couple to each other at a single-point in space --- we therefore call it the \emph{point-coupling Hamiltonian}. For a given frequency-independent classical scattering matrix implemented by the linear optical device, we provide a recipe to construct a point-coupling Hamiltonian describing the device. We formally integrate the Heisenberg equations of motion for the point-coupling Hamiltonian, and use the resulting solution to calculate its quantum scattering matrix \cite{fan2010input, xu2015input, trivedi2018few}. It is shown that an application of the quantum scattering matrix on an incoming quantum state is equivalent to applying the inverse of the classical scattering matrix on the annihilation operators of the optical modes in the incoming quantum state, thereby reproducing the commonly used procedure for analyzing the quantum physics of linear-optical devices. { We also `diagonalize' the point-coupling Hamiltonian to calculate its normal modes --- this provides a connection between the Hamiltonian presented in this paper and the quantization approach described in Ref.~\cite{dalton1999quasi}} \\

\noindent Finally, in order to demonstrate the utility of the point-coupling Hamiltonian proposed in this paper, we use it to rigorously derive and implement a matrix-product-state (MPS) \cite{schollwock2011density, orus2014practical} based update to simulate time-delayed feedback systems with linear optical devices providing feedback \cite{pichler2016photonic, grimsmo2015time}. Time-delayed feedback systems typically have a low-dimensional quantum system, such as a two-level system, coupling to a well-defined optical mode and the emission from the quantum system is used to re-excite the quantum system via a feedback path. The feedback path can be constructed using a linear-optical device such as a mirror, or using another quantum system. { Time-delayed feedback systems possibly provide a platform for generation of highly entangled states for quantum computation \cite{pichler2017universal, lubasch2018tensor}, quantum simulation \cite{dhand2018proposal} as well as implementation of quantum memories using long-lived bound states \cite{calajo2019exciting}.} MPS based simulations of such systems were proposed in Ref.~\cite{pichler2016photonic}, although the formulation relied on hard boundary conditions e.g.~it assumed that the feedback to the quantum system was provided by an ideal mirror. Using the point-coupling Hamiltonian, { we outline a method to extend the MPS based simulation of this system to account for the situation wherein the mirror is described by a scattering matrix, and thus enable an analysis of the impact of non-ideality in the mirror on the dynamics of the feedback-system.}  We expect the point-coupling Hamiltonian proposed in this paper to be of utility in analyzing quantum systems with complicated linear optical devices that can only be described by a full scattering matrix, and cannot be well-approximated by hard boundary conditions.\\

\noindent This paper is organized as follows --- Section \ref{sec:point_coupling_Hamiltonian} introduces the point-coupling Hamiltonian and analyzes it in the Heisenberg picture. Using the solution of the Heisenberg equations of motion, we explicitly calculate the quantum scattering matrix for the linear-optical device. It is shown that an application of the quantum scattering matrix on an incoming state is equivalent to applying the inverse of the classical scattering matrix on the annihilation operators in the state. We also provide the point-coupling Hamiltonians corresponding to some commonly used linear optical devices such as phase shifters, beam splitters and optical circulators. { Finally, we diagonalize the point-coupling Hamiltonian and show that the classical modes corresponding to the annihilation operators that diagonalize the Hamiltonian can be interpreted as the normal modes of the linear-optical device.} In section \ref{sec:mps}, we use the proposed point-coupling Hamiltonian to analyze feedback into a two-level system from a partially transmitting mirror using MPS update.
\section{Point-coupling Hamiltonian}\label{sec:point_coupling_Hamiltonian}
\subsection{Dynamics in the Heisenberg picture}
\begin{figure}[b]
\centering
\includegraphics[scale=0.65]{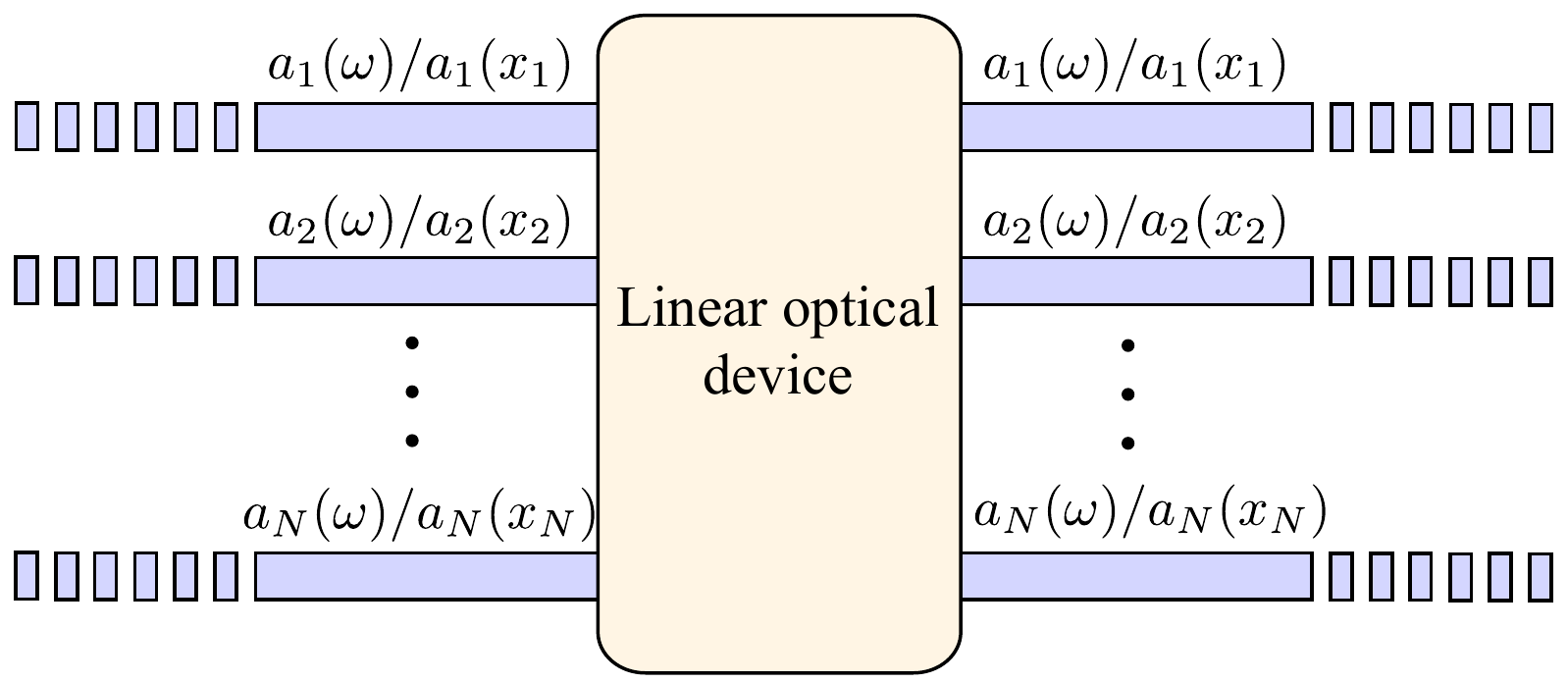}
\caption{Schematic of a linear optical device acting on $N$ optical modes. The frequency-domain and position-domain annihilation operator fo the $n^\text{th}$ optical mode are denoted by $a_n(\omega)$ and $a_n(x_n)$ respectively where $x_n$ is the position of the point under consideration in the coordinate system attached to the $n^\text{th}$ optical mode (note that we each optical mode to, in general, have its own independent coordinate system with the linear optical device being at $x_n = x_n^0$ in the coordinate system of the $n^\text{th}$ optical mode.).}
\label{fig:schematic_lin_opt}
\end{figure}

\noindent Consider $N$ propagating optical modes (which can physically be waveguide modes, or collimated optical beams) interacting with each other through a linear-optical device (Fig.~\ref{fig:schematic_lin_opt}). { Note that optical modes that are identical but for the direction of propagation are counted as separate modes.} Labelling by $a_n(\omega)$ the annihilation operator of the $n^\text{th}$ optical mode at frequency $\omega$, we propose the following Hamiltonian for describing the dynamics of the system:
\begin{align}\label{eq:point_coupling_Hamiltonian}
    H = \sum_{n=1}^N \int_{-\infty}^\infty \omega a_n^\dagger(\omega)a_n(\omega)\textrm{d}\omega +\sum_{n=1}^N\sum_{m=1}^N \text{V}_{n, m} a_n^\dagger (x_n^0) a_m(x_m^0),
\end{align}
where $a_n(x_n)$ is the position-domain annihilation operator for the $n^\text{th}$ optical mode at { coordinate $x_n$ along its direction of propagation}:
\begin{align}\label{eq:pos_annihilation_op}
    a_n(x_n) = \int_{-\infty}^\infty a_n(\omega)e^{\textrm{i}\omega x_n} \frac{\textrm{d}\omega}{\sqrt{2\pi}}.
\end{align}
{ Here the position $x_n$ is expressed in units of time such that the group velocity, assumed to be frequency independent and uniform across all the optical modes, is unity. The coordinate of the linear-optical device along the direction of propagation of the $n^\text{th}$ optical mode is $x_n^0$. Note that we allow for the coordinate $x_n$ for different optical modes to be expressed in different coordinate systems --- in general, each optical mode can have its own coordinate system with the $x$ axis of the coordinate system being along its direction of propagation.  For e.g.~the $x$ axis for a forward propagating mode will be in a direction opposite to the $x$ axis for a backward propagating mode, or modes waveguides oriented in physically different directions will have differently oriented coordinate systems. We note that $a_n(\omega)$, and consequently $a_n(x_n)$, satisfy the usual bosonic commutation relations: $[a_n(\omega), a_m^\dagger(\omega')] = \delta_{n, m}\delta(\omega - \omega')$, $[a_n(\omega), a_m(\omega')] = 0$, $[a_n(x_n), a_m^\dagger(x'_m)] = \delta_{n, m}\delta(x_n - x'_n)$ and $[a_n(x_n), a_m(x'_m)] = 0$. Moreover, as a consequence of the Hermiticity of $H$, the coupling coefficients $\text{V}_{k, m}$ satisfy $\text{V}_{n, m} = \text{V}_{m, n}^*$.}\\ \ \\
\noindent The dynamics of this Hamiltonian can be easily analyzed using the Heisenberg's equations of motion, which are given by:
\begin{align}\label{eq:heisenberg_eq}
    \textrm{i}\frac{\textrm{d}a_n(\omega; t)}{\textrm{d}t} = \omega a_n(\omega; t) +\sum_{m=1}^N  \frac{\text{V}_{n, m} }{\sqrt{2\pi}}a_m(x_m^0; t) e^{-\textrm{i}\omega x_n^0} \ \forall \ n \in \{1, 2 \dots N\}.
\end{align}
As is shown in appendix \ref{app:heisenberg}, these equations can easily be integrated to relate the position-domain annihilation operators at time $t = t_0 + \tau$ and displaced by a distance $x \in (-\infty, \infty)$ from the linear optical device to the position-domain annihilation operators at time $t = t_0 $:
\begin{align}\label{eq:sol_heisenberg}
    \begin{bmatrix}
    a_1(x_1^0 + x; t_0 + \tau) \\
    a_2(x_2^0 + x; t_0 + \tau) \\
    \vdots \\
    a_N(x_N^0 + x; t_0 + \tau)
    \end{bmatrix} = 
    \bigg[\textbf{I}-\textrm{i}\textbf{V}\bigg(\textbf{I} + \frac{\textrm{i}\textbf{V}}{2}\bigg)^{-1} \Theta(0\leq y \leq \tau)\bigg]\begin{bmatrix}
    a_1(x_1^0 + x - \tau; t_0) \\
    a_2(x_2^0 + x - \tau; t_0) \\
    \vdots \\
    a_N(x_N^0 + x - \tau; t_0)
    \end{bmatrix},
\end{align}
where $\textbf{V}$ is a $N \times N$ Hermitian matrix formed by $\text{V}_{m, n}$ as its elements, $\textbf{I}$ is the identity matrix of size $N$ and $\Theta(x_1 \leq x \leq x_2)$ is defined by:
\begin{align}\label{eq:step_fun}
    \Theta(x_1 \leq x \leq x_2) = 
    \begin{cases}
    1 & \text{if } x \in (x_1, x_2) \\
    0 & \text{if } x \in (-\infty, x_1) \cup (x_2, \infty)\\
    \frac{1}{2} & \text{if } x \in \{x_1, x_2\}
    \end{cases}.
\end{align}
To intuitively interpret the result in Eq.~\ref{eq:sol_heisenberg}, note that if $y< 0$, then the position-domain annihilation operator at $t = t_0 + \tau$ is simply a propagated version of itself at $t = t_0$. This is a simply a consequence of the physical points described by $x < 0$ lying before the linear optical device along the propagation direction, and consequently being unaffected by scattering from the linear-optical device. This situtation is the same for $x >  \tau$, since the scattered light from the linear-optical device has not had sufficient time to propagate to the points in question from the location of the linear optical device (i.e.~from $x = 0$). For $0 < x < \tau $, the optical mode annihilation operator at $t = t_0 + \tau$ is a sum of a propagated version of itself and contributions from other optical modes as scattered by the linear optical device, and Eq.~\ref{eq:sol_heisenberg} can be simplified to:
\begin{align}
    \begin{bmatrix}
    a_1(x_1^0 + x; t_0 + \tau) \\
    a_2(x_2^0 + x; t_0 + \tau) \\
    \vdots \\
    a_N(x_N^0 + x; t_0 + \tau)
    \end{bmatrix} = 
    \textbf{S}
    \begin{bmatrix}
    a_1(x_1^0 + x - \tau; t_0) \\
    a_2(x_2^0 + x - \tau; t_0) \\
    \vdots \\
    a_N(x_N^0 + x - \tau; t_0)
    \end{bmatrix},
\end{align}
where
\begin{align}\label{eq:classical_scat_mat}
    \textbf{S} = \textbf{I} - \textrm{i}\textbf{V}\bigg( \textbf{I} + \frac{\textrm{i}\textbf{V}}{2}\bigg)^{-1} = \bigg(\textbf{I} - \frac{\textrm{i}\textbf{V}}{2} \bigg)\bigg(\textbf{I} + \frac{\textrm{i}\textbf{V}}{2} \bigg)^{-1}.
\end{align}
Since $\textbf{S}$ relates the optical mode annihilation operators before and after scattering from the linear optical-device has occurred, it can be interpreted as the classical scattering matrix corresponding to the linear optical-device. It can readily be verified that a consequence of $\textbf{V}$ being Hermitian is that the matrix $\textbf{S}$ is unitary. Moreover, if $\textbf{S}$ has the diagonalization $\textbf{S} = \textbf{U}\ \text{diag}\big[e^{\textrm{i}\boldsymbol{\phi}}\big]\ \textbf{U}^\dagger$, it follows from Eq.~\ref{eq:classical_scat_mat} that:
\begin{align}\label{eq:Hamiltonian_classical_scat_mat}
\textbf{V} = -\textbf{U}\ \text{diag}\bigg[2\tan\bigg(\frac{\boldsymbol{\phi}}{2}\bigg)\bigg]\ \textbf{U}^\dagger.
\end{align}
Therefore, given the classical matrix $\textbf{S}$ of a linear-optical device, Eq.~\ref{eq:Hamiltonian_classical_scat_mat} allows us to construct the matrix $\textbf{V}$ and by extension the point-coupling Hamiltonian that can model the device. As examples, we consider some commonly used linear optical devices (Fig.~\ref{fig:schematic_examples}) and construct the point-coupling Hamiltonians that describe their dynamics:
\begin{figure}[b]
\centering
\includegraphics[scale=0.4]{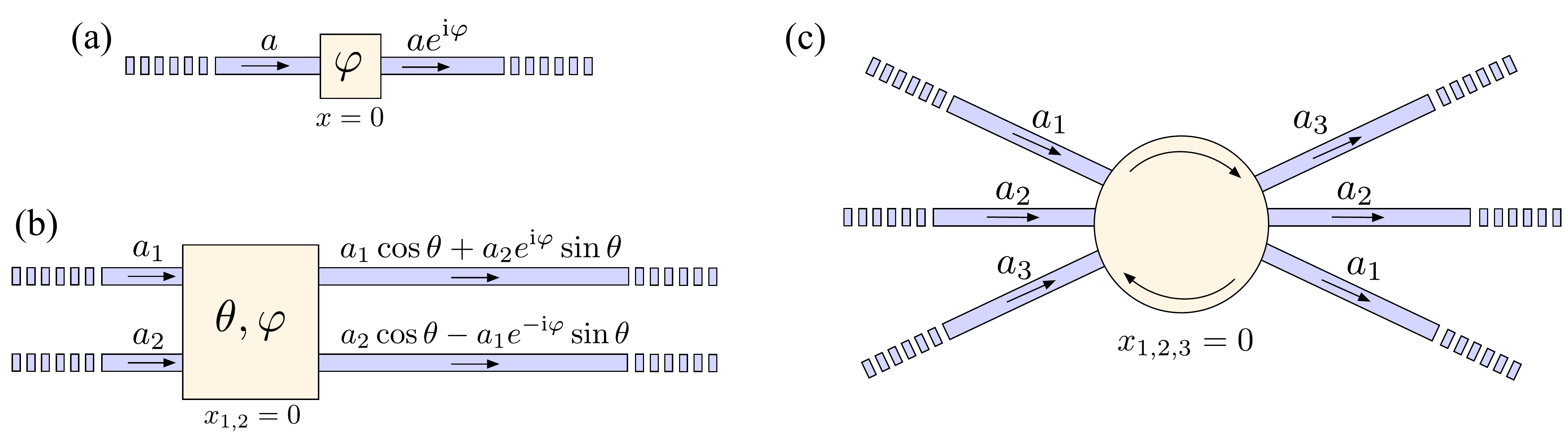}
\caption{Schematic figures showing the action of (a) phase-shifter with imparted phase $\varphi$ acting on a optical mode denoted by $a$, (b) beam-splitter with parameters $(\theta, \varphi)$ acting on optical modes denoted by $a_1$ and $a_2$ and (c) an optical circulator designed acting on optical modes $a_1$, $a_2$ and $a_3$. The relationship between the input and output signals of the linear-optical device are also shown in the diagrams.}
\label{fig:schematic_examples}
\end{figure}

\begin{enumerate}
\item[(a)] \emph{Phase shifter}: The classical scattering matrix $\textbf{S}$ of the phase-shifter [Fig.~\ref{fig:schematic_examples}(a)]  is a single-element matrix: $\textbf{S} = [e^{\textrm{i}\varphi}]$. From Eq.~\ref{eq:Hamiltonian_classical_scat_mat}, we obtain $\textbf{V} = [-2\tan(\varphi / 2)]$. The Hamiltonian for the phase shifter can thus be written as:
\begin{align}
H_\text{phase-shifter} = \int_{-\infty}^\infty \omega a^\dagger(\omega) a(\omega)\textrm{d}\omega - 2 \tan\bigg(\frac{\varphi}{2}\bigg) a^\dagger(x = 0) a(x=0),
\end{align}
where $a(\omega)$ [$a(x)$] is the frequency-domain (position-domain) annihilation operator for the optical mode that the phase-shifter is acting on.\\

\item[(b)] \emph{Beam-splitter}: The beam-splitter [Fig.~\ref{fig:schematic_examples}(b)] is described by the following classical scattering matrix:
\begin{align}
\textbf{S} = \begin{bmatrix}
\cos \theta & \sin \theta \ e^{\textrm{i}\varphi} \\
-\sin \theta \ e^{-\textrm{i}\varphi} & \cos \theta 
\end{bmatrix}.
\end{align}
Again, using Eq.~\ref{eq:Hamiltonian_classical_scat_mat}, we obtain
\begin{align}
\textbf{V} = \begin{bmatrix}
0 & 2\textrm{i}\tan (\theta / 2)\ e^{\textrm{i}\varphi} \\
-2\textrm{i}\tan(\theta / 2)\ e^{-\textrm{i}\varphi} & 0
\end{bmatrix},
\end{align}
from which we can construct the beam splitter Hamiltonian:
\begin{align}
H_\text{beam-splitter} = \sum_{k \in \{1, 2\}} \int_{-\infty}^\infty \omega a_k^\dagger(\omega) a_k(\omega)\textrm{d}\omega + \bigg[2\textrm{i}\tan\bigg(\frac{\theta}{2}\bigg)e^{\textrm{i}\varphi}a_1^\dagger(x_1 = 0)a_2(x_2 = 0)+\textrm{h.c.} \bigg],
\end{align}
where $a_{1,2}(\omega)$ [$a_{1,2}(x_{1, 2})$] are the frequency-domain (position-domain) annihilation operators for the optical modes that the beam-splitter is acting on.\\

\item[(c)] \emph{Optical circulator}: The optical circulator [Fig.~\ref{fig:schematic_examples}(c)] routes an excitation in optical mode 1 to optical mode 2, optical mode 2 to optical mode 3 and optical mode 3 to optical mode 1. It can thus be described by the following $3\times 3$ classical scattering matrix:
\begin{align}
\textbf{S} = \begin{bmatrix}
0 & 1 & 0 \\
0 & 0 & 1 \\
1 & 0 & 0
\end{bmatrix},
\end{align}
and therefore
\begin{align}
\textbf{V} = \begin{bmatrix}
0 & 2 \textrm{i} & -2\textrm{i} \\
-2\textrm{i} & 0 & 2 \textrm{i} \\
2\textrm{i} & -2\textrm{i} & 0
\end{bmatrix},
\end{align}
from which we can construct the optical circulator Hamiltonian:
\begin{align}
H_\text{circulator} = \sum_{k \in \{1, 2, 3\}} \int_{-\infty}^\infty \omega a_k^\dagger(\omega) a_k(\omega) \textrm{d}\omega + \bigg[\sum_{k \in \{1, 2, 3\}} 2\textrm{i}a_k^\dagger(x_k = 0) a_{k + 1}(x_{k + 1} = 0) +\textrm{h.c.}\bigg],
\end{align}
where $a_{1, 2, 3}(\omega)$ [$a_{1, 2, 3}(x_{1, 2, 3})$] are the frequency-domain (position-domain) annihilation operators for the optical modes that the optical circulator acts on, and $a_4(\omega) / a_4(x_4)$ is to be interpreted as $a_1(\omega) / a_1(x_1)$.
\end{enumerate}

\subsection{Quantum scattering matrix of the point-coupling Hamiltonian}
\noindent While the previous subsection analyzed the point-coupling Hamiltonian in the Heisenberg picture, in this subsection we analyze its dynamics in the Schroedinger picture. In particular, we consider the problem of exciting the linear-optical device with an input state, and attempt to calculate the output state produced by the device.  \\ \ \\
\noindent The key object relating the input state (assumed to be the asymptote \cite{taylor2006scattering} to the system at $t \to -\infty$) of the system to its output state (assumed to be the asymptote \cite{taylor2006scattering} to the state of the system at $t \to \infty$)  is the quantum \cite{taylor2006scattering}:
\begin{align}
    \mathcal{S} = \lim_{\substack{t_+ \to \infty \\ t_- \to -\infty}} e^{\textrm{i}H_0 t_+}e^{-\textrm{i}H(t_+ - t_-)}e^{-\textrm{i}H_0 t_-},
\end{align}
where $H_0$ is the Hamiltonian of the optical modes without accounting for the linear-optical device:
\begin{align}
    H_0 = \sum_{k=1}^N \int_{-\infty}^\infty \omega a_k^\dagger(\omega) a_k(\omega) \textrm{d}\omega.
\end{align}
Consider now the computation of the following $K$ photon matrix element of the scattering matrix ($\textbf{x} \equiv \{x_1, x_2 \dots x_K\}$, $\textbf{x}' \equiv \{x_1', x_2' \dots x_K'\}$, $\boldsymbol{\mu} = \{\mu_1, \mu_2 \dots \mu_K\}$ and $\boldsymbol{\mu}' = \{\mu_1', \mu_2' \dots \mu_K'\}$):
\begin{align}\label{eq:smat_elem}
\mathcal{S}(\textbf{x}, \boldsymbol{\mu}; \textbf{x}', \boldsymbol{\mu}') = \bra{\text{vac}} \bigg[\prod_{i=1}^K a_{\mu_i}(x_i)\bigg] \mathcal{S} \bigg[\prod_{i=1}^K a_{\mu_i'}^\dagger(x_i') \bigg]\ket{\text{vac}}.
\end{align}
Note that the relations $\exp(-\textrm{i}H_0 t) a_k(x_k)\exp(\textrm{i}H_0 t) = a_k(x_k + t)$ and $\exp(\textrm{i} H t) a_k(x_k) \exp(-\textrm{i}H t) = a_k(x_k; t)$ together with $H_0 \ket{\text{vac}} = H\ket{\text{vac}} = 0$ immediately imply the following relationship between the scattering matrix element in Eq.~\ref{eq:smat_elem} and the Heisenberg picture optical mode position-domain annihilation operators:
\begin{align}\label{eq:smat_elem_heisenberg}
    \mathcal{S}(\textbf{x}, \boldsymbol{\mu}; \textbf{x}', \boldsymbol{\mu}') = \lim_{\substack{t_+ \to \infty \\ t_- \to -\infty}}\bra{\text{vac}} \bigg[\prod_{i=1}^K a_{\mu_i}(x_i + t_+; t_+)\bigg]
    \bigg[\prod_{i=1}^K a_{\mu_i'}^\dagger(x_i' + t_-; t_-)\bigg] \ket{\text{vac}}.
\end{align}
Using Eq.~\ref{eq:sol_heisenberg}, and in the limit of $t_+ \to \infty$ and $t_- \to -\infty$, it follows that :
\begin{align}
a_{\mu_i}(x_i + t_+; t_+) = \sum_{\nu_i=1}^N \text{S}_{\mu_i, \nu_i} a_{\nu_i}(x_i - x_{\mu_i}^0 + x_{\nu_i}^0 + t_-; t_-),
\end{align}
where $S_{i, j}$ are the elements of the classical scattering matrix $\textbf{S}$. { With this, the following explicit expression for $\mathcal{S}(\textbf{x}, \boldsymbol{\mu}; \textbf{x}', \boldsymbol{\mu}')$ as given by Eq.~\ref{eq:smat_elem_heisenberg} can be obtained:
\begin{align}\label{eq:pos_dom_scat_mat}
\mathcal{S}(\textbf{x}, \boldsymbol{\mu}; \textbf{x}', \boldsymbol{\mu}') = \sum_{\mathcal{P}_{K}} \prod_{l=1}^K \text{S}_{\mu_l, \mathcal{P}_K\mu'_l} \delta\big((x_l - x_{\mu_l}^0) - (\mathcal{P}_K x_l' - x^0_{\mathcal{P}_{K} \mu'_l})\big),
\end{align}
where $\mathcal{P}_K$ is a $K$ element permutation. It can immediately be noticed that the quantum scattering matrix elements are completely determined in terms of the classical scattering matrix elements $\text{S}_{i, j}$. From Eq.~\ref{eq:pos_dom_scat_mat}, we can also evaluate the frequency domain scattering matrix elements $S(\boldsymbol{\omega}, \boldsymbol{\mu}; \boldsymbol{\omega}', \boldsymbol{\mu}')$ ($\boldsymbol{\omega} = \{\omega_1, \omega_2 \dots \omega_K\}$, $\boldsymbol{\omega}' = \{\omega_1', \omega_2' \dots \omega_K'\}$, $\boldsymbol{\mu} = \{\mu_1, \mu_2 \dots \mu_K\}$ and $\boldsymbol{\mu}' = \{\mu_1', \mu_2' \dots \mu_K'\}$):
\begin{align}\label{eq:freq_dom_scat_mat}
\mathcal{S}(\boldsymbol{\omega}, \boldsymbol{\mu}; \boldsymbol{\omega}', \boldsymbol{\mu}') = \int_{\mathbb{R}^K} \int_{\mathbb{R}^K} S(\textbf{x}, \boldsymbol{\mu}; \textbf{x}', \boldsymbol{\mu}') \bigg[\prod_{l=1}^K \frac{e^{\textrm{i}(\omega_l' x_l' - \omega_l x_l)}}{2\pi}\bigg]\textrm{d}^K\textbf{x}\ \textrm{d}^K\textbf{x}' =\sum_{\mathcal{P}_K} \prod_{l=1}^K \textrm{S}_{ \mathcal{P}_K \mu_l,\mu_l'}\ e^{-\textrm{i}\omega_l \big(x^0_{\mathcal{P}_K\mu_l} - x^0_{ \mu_l'}\big)} \delta(\mathcal{P}_K \omega_l - \omega_l').
\end{align}
Note that the frequency domain scattering matrix doesn't have any connected parts \cite{xu2015input} i.e.~scattering of a $K$ photon wave-packet from the linear optical device conserves the individual input frequencies. This is a direct consequence of the `linearity' of the optical device. To gain more insight into the form of the scattering matrix in Eqs.~\ref{eq:pos_dom_scat_mat} and \ref{eq:freq_dom_scat_mat}, consider the calculation of the output state corresponding to a $K$-photon input state $\ket{\psi_\text{in}}$:
\begin{align}\label{eq:K_ph_input_state}
\ket{\psi_\text{in}} = \sum_{\boldsymbol{\mu}} \int_{\mathbb{R}^K} \psi_\text{in}(\boldsymbol{\omega}, \boldsymbol{\mu}) \bigg[\prod_{l=1}^K a_{\mu_l}^\dagger(\omega_l)\bigg] \ket{\text{vac}} \textrm{d}^K \boldsymbol{\omega},
\end{align}
where the amplitude $\psi_\text{in}(\boldsymbol{\omega}, \boldsymbol{\mu})$ can be chosen, without any loss of generality, to be symmetric with respect to a simultaneous permutation of the indices $\boldsymbol{\mu}$ and $\boldsymbol{\omega}$: $\psi_\text{in}(\boldsymbol{\omega}, \boldsymbol{\mu}) = \psi_\text{in}(\mathcal{P}_K \boldsymbol{\omega}, \mathcal{P}_K\boldsymbol{\mu}) \ \forall \ K$-element permutations $\mathcal{P}_K$. The output state is then given by:
\begin{align}\label{eq:K_ph_output_state}
\ket{\psi_\text{out}} = \sum_{\boldsymbol{\mu}} \int_{\mathbb{R}^K} \psi_\text{out}(\boldsymbol{\omega}, \boldsymbol{\mu}) \bigg[\prod_{l=1}^K a_{\mu_l}^\dagger(\omega_l)\bigg] \ket{\text{vac}} \textrm{d}^K \boldsymbol{\omega},
\end{align}
where $\psi_\text{out}(\boldsymbol{\omega}, \boldsymbol{\mu})$ is given by:
\begin{align}\label{eq:input_output_state}
\psi_\text{out}(\boldsymbol{\omega}, \boldsymbol{\mu}) = \frac{1}{N!} \sum_{\boldsymbol{\mu}'}\int_{\mathbb{R}^K} S(\boldsymbol{\omega}, \boldsymbol{\mu}; \boldsymbol{\omega}', \boldsymbol{\mu}')\psi_\text{in}(\boldsymbol{\omega}', \boldsymbol{\mu}')\textrm{d}^K \boldsymbol{\omega}'.
\end{align}
Assuming that the coordinate systems of the optical modes are chosen such that $x_k^0 = 0 \ \forall \ k \in \{1, 2 \dots N\}$, from Eqs.~\ref{eq:freq_dom_scat_mat} and \ref{eq:input_output_state} we obtain:
\begin{align}
\psi_\text{out}(\boldsymbol{\omega}, \boldsymbol{\mu}) &= \frac{1}{N!} \sum_{\boldsymbol{\mu}'}\sum_{\mathcal{P}_K} \bigg[\prod_{l=1}^K \textrm{S}_{ \mu_l, \mathcal{P}_K\mu_l'} \bigg] \psi_\text{in}( \boldsymbol{\omega}, \mathcal{P}_K\boldsymbol{\mu}').
\end{align}
Therefore,
\begin{align}
\ket{\psi_\text{out}} &=\sum_{\boldsymbol{\mu}} \int_{\mathbb{R}^K} \psi_\text{in}(\boldsymbol{\omega}, \boldsymbol{\mu}) \bigg[\prod_{l=1}^K \tilde{a}^\dagger_{\mu_l}(\omega_l) \bigg] \ket{\text{vac}}\ \textrm{d}^K \boldsymbol{\omega},
\end{align}
where $\tilde{a}_i(\omega) = \sum_{i=1}^N \text{S}_{j, i}^* a_j(\omega)$}. The application of the quantum scattering matrix of a linear optical device to an input state is equivalent to replacing the annihilation operators in the input state with $\textbf{S}^\dagger$ times the annihilation operators. This is the usual procedure used to analyze the impact of a linear optical element on an input state \cite{mandel1995optical, kok2007linear}, and our analysis derives it from a Hamiltonian based description of the linear optical element and thus lends rigor to this procedure.

{ \subsection{Normal modes of the point-coupling Hamiltonian}\label{sec:normal_modes}
In this subsection, we diagonalize the point-coupling Hamiltonian and explicitly calculate the normal modes of the linear-optical device. Given that the frequency-domain annihilation operator of the $n^\text{th}$ normal mode is $b_n(\omega)$, it should satisfy the following commutation relations:
\begin{subequations}\label{eq:diag_cond}
\begin{align}
&[b_n(\omega), H] = \omega b_n(\omega), \label{eq:comm_with_hamil}\\
&[b_n(\omega), b_m^\dagger(\omega')] = \delta_{n, m}\delta(\omega - \omega'), \label{eq:comm_with_itself}
\end{align}
\end{subequations}
where $H$ is the point-coupling Hamiltonian. Eq.~\ref{eq:comm_with_hamil} is a consequence of the Hamiltonian $H$ being expressible as a sum of independent continua of harmonic oscillators in the normal mode basis:
\begin{align}
H = \sum_{n =1}^N \int_{-\infty}^\infty \omega b_n^\dagger(\omega) b_n(\omega) \textrm{d}\omega,
\end{align}
and Eq.~\ref{eq:comm_with_itself} is a result of the different normal modes being physically independent modes. As is shown in appendix \ref{app:diagonalization_details}, using Eqs.~\ref{eq:comm_with_hamil} and \ref{eq:comm_with_itself}, the normal mode annihilation operators $b_n(\omega)$ can be chosen to be:
\begin{align}\label{eq:diag_result}
b_n(\omega) =\int_{-\infty}^{x_n^0} e^{-\textrm{i}\omega (x-x_n^0)} a_n(x_n) \frac{\textrm{d}x_n}{\sqrt{2\pi}} + \sum_{m=1}^N \text{S}_{m, n}^*\int_{x_m^0}^\infty e^{-\textrm{i}\omega (x-x_m^0)} a_m(x_m) \frac{\textrm{d}x_m}{\sqrt{2\pi}},
\end{align}
where $\text{S}_{n, m}$ are the elements of the classical scattering matrix. To obtain a physical interpretation of this result, consider creating a photon in the normal mode described by $b_n(\omega)$ and calculating its projection on the modes described by $a_m(x_m)$ --- this projection is given by the expectation $\bra{\text{vac}}a_m(x_m)b_n^\dagger(\omega)\ket{\text{vac}}$ which can be readily evaluated using Eq.~\ref{eq:diag_result}:
\begin{align}
\bra{\text{vac}}a_m(x_m) b_n^\dagger(\omega)\ket{\text{vac}} = \begin{cases}
e^{\textrm{i}\omega (x_m - x_m^0)} \delta_{m, n} & \text{if } x_m < x_m^0 \\
e^{\textrm{i}\omega (x_m - x_m^0)} \text{S}_{m, n}  & \text{if } x_m > x_m^0
\end{cases}.
\end{align}
This is exactly the same field profile that would be obtained on classically exciting the linear optical device through the $n^\text{th}$ input port at frequency $\omega$, and calculating the fields scattered in all the output ports --- the $n^\text{th}$ normal mode is simply the continuum of quantum harmonic oscillators associated with this field profile.\\ \ \\
Finally, Eq.~\ref{eq:diag_result} can be inverted to relate the annihilation operators $a_n(x)$ to the normal mode annihilation operators $b_n(\omega)$ (refer to appendix \ref{app:diagonalization_details} for details):
\begin{align}\label{eq:diag_inv_result}
a_n(x_n) = \int_{-\infty}^{\infty} \frac{\textrm{d}\omega  }{\sqrt{2\pi}}e^{\textrm{i}\omega (x_n - x_n^0)} \begin{cases}
b_n(\omega) & \text{if } x_n > x_n^0\\
\sum_{m=1}^N \text{S}_{n, m} b_m(\omega) & \text{if } x_n < x_n^0
\end{cases},
\end{align}
where, again, we see that the modes described by $a_n(x)$ are identical to the normal modes at points before the linear-optical device (i.e. $x_n < x_n^0$ in Eq.~\ref{eq:diag_inv_result}) and their linear combination at points after the linear-optical device. Eqs.~\ref{eq:diag_result} and \ref{eq:diag_inv_result} thus provides a connection between the point-coupling Hamiltonian for linear-optical device and a direct quantization of the normal modes of the device \cite{dalton1999quasi}.
}

\section{Matrix-product-state based simulations of time-delayed feedback systems}\label{sec:mps}
\begin{figure}[b]
\centering
\includegraphics[scale=0.4]{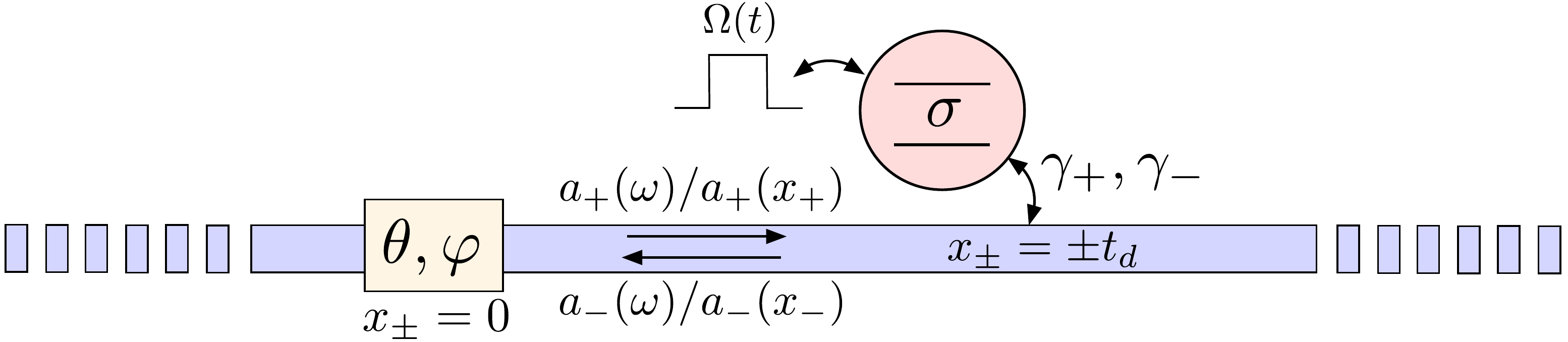}
\caption{Schematic figure of the time-delayed feedback system analyzed in this paper. A two-level system with de-excitation operator $\sigma$ couples to a waveguide which supports both forward propagating and backward propagating waveguide modes with frequeny-domain (position-domain) annihilation operators $a_+(\omega)$ and $a_-(\omega)$ ($a_+(x_+)$ and $a_-(x_-)$) respectively. A mirror, modelled as a beam-splitter on the forward and backward propagating waveguide mode with parameters $(\theta, \varphi)$, is located as $x_\pm = 0$ and the two-level system couples to both the waveguide modes at a distance $t_d$ from the mirror. The decay rates of the two-level system into the forward and backward propagating waveguide modes are denoted by $\gamma_+$ and $\gamma_-$ respectively. Finally, we also consider the dynamics of this system when the two-level system is driven by an external laser pulse --- $\Omega(t)$ denotes the complex amplitude of the laser pulse that drives the two-level system.}
\label{fig:schematic_tls}
\end{figure}
\noindent In this section, we use the point-coupling Hamiltonian proposed in the section \ref{sec:point_coupling_Hamiltonian} to develop an MPS update for a two-level system coupled to a waveguide with a partially transmitting mirror. Using the formulated MPS update, we study the impact of the less than unity reflectivity of the mirror on the dynamics of the time-delayed feedback system.

\begin{figure}[t]
\centering
\includegraphics[scale=0.36]{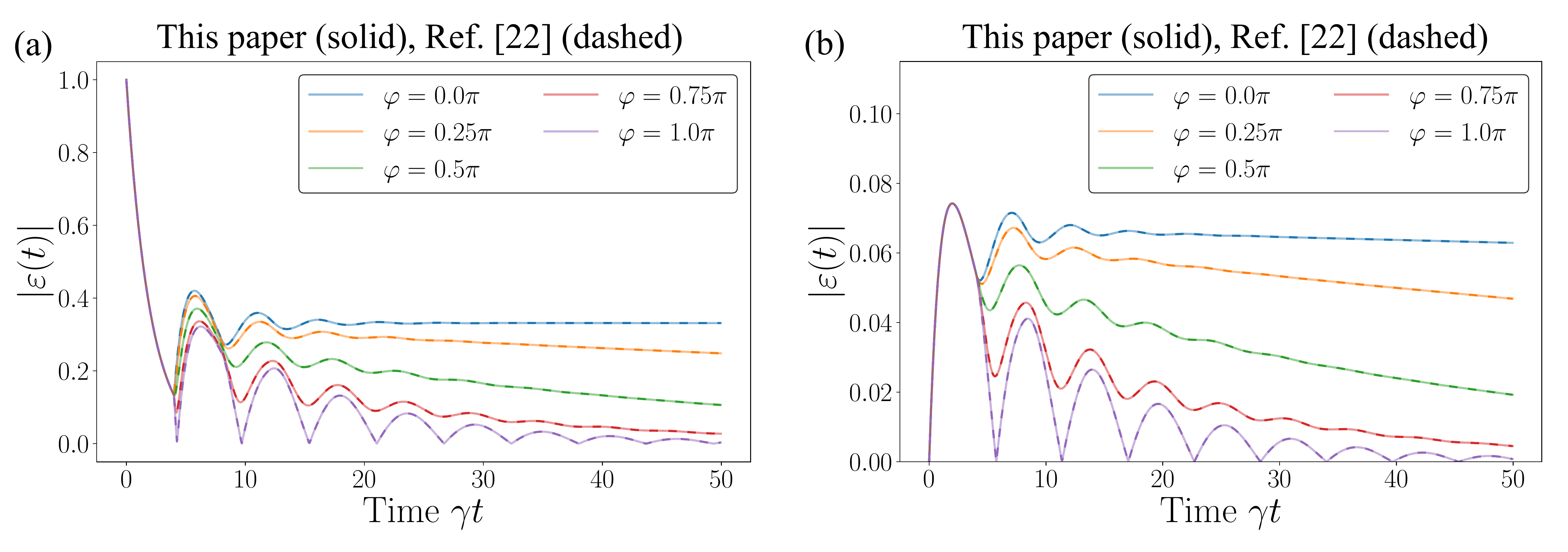}
\caption{Validation of our MPS update implementation against the implementation introduced in Ref.~\cite{pichler2016photonic} for an ideal mirror (i.e.~$\theta = \pi / 2$) for two simulation settings: (a) Simulation of an undriven two-level system ($\Omega(t) = 0$) which is initially in its excited state for different mirror phases $\varphi$ and (b) simulation of a two-level system initially in its ground state and driven by an exponentially decaying pulse ($\Omega(t) = \Omega_0 e^{-\alpha t}$ for $t > 0$) for different mirror phases $\varphi$. $|\varepsilon(t)|$ is the probability amplitude of the two-level system being in the excited state computed using $|\varepsilon(t)|^2 = \langle \sigma^\dagger \sigma \rangle$. It is assumed that $\gamma_+ = \gamma_- = \gamma / 2$, $\delta_e = \omega_e - \omega_0 = 0$, $\omega_0 t_d = \pi$, $\gamma t_d = 2$ and $\alpha = 2\gamma$. For the discretization into an MPS, we use $\gamma\Delta t = 0.05$, and truncate the dimensionality of the Hilbert space of each waveguide bin to 2 for both forward and backward propagating modes. A threshold of $0.01$ is used in all the Schmidt decompositions performed while applying the swap gates and the short-range gates. { Refer to appendix \ref{app:mps_conv} for convergence studies of the MPS simulations.}}
\label{fig:validation_fig}
\end{figure}

\noindent The system under consideration is a two-level system with a time-delayed feedback shown in Fig.~\ref{fig:schematic_tls}. The Hamiltonian for this system can be expressed as a sum of a two-level system Hamiltonian, waveguide Hamiltonian and the mirror Hamiltonian (which is modeled as a point interaction between the forward and backward propagating modes):
\begin{align}
    H(t) = H_\text{TLS}(t) + H_\text{wg} + H_\text{mirror} + H_\text{wg-TLS},
\end{align}
where $H_\text{TLS}(t)$ is the Hamiltonian of the two-level system including a coherent drive, $H_\text{wg}$ is the Hamiltonian describing the forward and backward propagating waveguide modes, $H_\text{mirror}$ is the Hamiltonian of the mirror providing feedback to the two-level system and $H_\text{wg-TLS}$ is the interaction Hamiltonian between the two-level system and the two waveguide modes:
\begin{subequations}
\begin{align}
    &H_\text{TLS}(t) = \omega_e \sigma^\dagger \sigma + \Omega(t) \big(\sigma e^{\textrm{i}\omega_0 t} + \sigma^\dagger e^{-\textrm{i}\omega_0 t}\big), \\ 
    &H_\text{wg} = \sum_{k \in \{+, -\}} \int_{-\infty}^\infty \omega a_k^\dagger(\omega) a_k(\omega) \textrm{d}\omega, \\
    &H_\text{mirror} =  2\textrm{i}\tan\bigg(\frac{\theta}{2} \bigg)\big[ e^{\textrm{i}\varphi}a_+(x_+ = 0)a_-^\dagger(x_- = 0) - e^{-\textrm{i}\varphi} a_-(x_- = 0) a_+^\dagger(x_+ = 0)   \big],\\
    &H_\text{wg-TLS} = \big[\sqrt{\gamma_+} \sigma^\dagger a_+(x_+ = t_d) + \sqrt{\gamma_-}\sigma^\dagger a_-(x_- = -t_d) + \text{h.c.}\big].\label{eq:wg_tls_hamil}
   \end{align}
\end{subequations}
Here $\sigma$ is the de-excitation operator for a two-level system with resonant frequency $\omega_e$, $a_+(\omega)$ and $a_-(\omega)$ are the frequency-domain annihilation operators for the forward and backward propagating waveguide modes respectively and $a_+(x_+)$ and $a_-(x_-)$ are the position-domain annihilation operators for the forward and backward propagating waveguide modes respectively. Note that the $x$-axis for the coordinate systems for the forward and backward propagating waveguide modes are chosen to be in their direction of propagation with the same origin. A mirror, modeled with a beam splitter Hamiltonian between the forward and backward propagating waveguide modes with parameters $(\theta, \varphi)$, is located at $x_\pm = 0$. The two-level system interacts with the waveguide (with a decay rate $\gamma_+$ into the forward propagating waveguide mode and $\gamma_-$ into the backward propagating waveguide mode) at a distance of $t_d$ from the mirror --- this corresponds to the point with coordinates $x_+ = t_d$ and $x_- = -t_d$ in the coordinate systems of the forward and backward propagating waveguide modes. Furthermore, the two-level system is driven with a laser pulse at frequency $\omega_0$ and pulse-shape $\Omega(t)$.\\ \ \\
{ \noindent We first go into a frame rotating as per the Hamiltonian $H_0 = \omega_0 \sigma^\dagger \sigma + H_\text{wg} + H_\text{mirror}$. In this frame, the state of the system evolves as per the Hamiltonian $\tilde{H}(t)$:
\begin{align}\label{eq:rot_hamil}
\tilde{H}(t) = \delta_e \sigma^\dagger \sigma + \Omega(t)\sigma_x + \big(\sqrt{\gamma_+}\sigma^\dagger a_+(x_+ = t_d; t) e^{\textrm{i}\omega_0 t} + \sqrt{\gamma_-}\sigma^\dagger a_-(x_- = -t_d; t) e^{\textrm{i}\omega_0 t} + \text{h.c.}\big),
\end{align}
where $\delta_e = \omega_e - \omega_0$ and $a_\pm(x_\pm = \pm t_d; t)$ are the Heisenberg picture operators corresponding to $a_\pm(x_\pm=\pm t_d)$ with respect to $H_0$ at time $t$, subject to the initial condition $a_\pm(x_\pm = \pm t_d; t = 0) = a_\pm(x_\pm = \pm t_d)$. From Eq.~\ref{eq:sol_heisenberg} and for $t > 0$, we obtain:
\begin{subequations}\label{eq:rotating_frame_ref_op}
\begin{align}
&a_+(x_+ = t_d; t) = e^{-\textrm{i}\omega_0 ( t - t_d)}\begin{cases}
A_+(t) & \text{if } t < t_d \\
A_+(t) \cos \theta + e^{\textrm{i}\varphi} A_-(t - 2t_d) \sin \theta & \text{if } t > t_d
\end{cases},\\
&a_-(x_- = -t_d; t) = e^{-\textrm{i}\omega_0 (t + t_d)} A_-(t),
\end{align}
\end{subequations}
where we have defined the operators $A_\pm(t)$ via:
\begin{align}
A_\pm(t) = \int_{-\infty}^{\infty} a_\pm(\omega) e^{-\textrm{i}(\omega - \omega_0)(t \mp t_d)} \frac{\textrm{d}\omega}{\sqrt{2\pi}}.
\end{align}
An MPS update for the state of the system in the rotating frame with respect to $H_0$ can now be framed using the procedure outlined in Ref.~\cite{pichler2016photonic} --- the first step is to discretize the waveguide Hilbert space. Using a discretization step $\Delta t$, we can define the waveguide bin operators $A_\pm[k]$ in terms of the operators $A_\pm(t)$ via:
\begin{align}\label{eq:disc}
A_\pm[k] = \int_{k\Delta t}^{(k + 1)\Delta t} A_\pm(t) \frac{\textrm{d} t}{\sqrt{\Delta t}},
\end{align}}
which satisfy the commutation relations $[A_+[i], A_+^\dagger[j]] = \delta_{i, j}$ and $[A_-[i], A_-^\dagger[j]] = \delta_{i, j}$. The state of the entire system, including the waveguide and the two-level system, can be represented by a matrix-product state with the two-level system corresponding to the first site in the matrix product state, and the subsequent sites corresponding to the waveguide bins. The Hilbert space of the $k^\text{th}$ waveguide bin is spanned by the tensor product of the Hilbert spaces of the harmonic oscillators whose annihilation operators are $A_+[k]$ and $A_-[k]$. To compute the the state at $t = k\Delta t$ from the state at $t = (k + 1)\Delta t$, we act on the matrix product state with the unitary operator $U[k + 1, k]$ defined by:
\begin{align}
U[k + 1, k] = e^{-\textrm{i} H[k + 1, k]},
\end{align}
where
\begin{align}
H[k + 1, k] &= \int_{k\Delta t}^{(k + 1) \Delta t} \tilde{H}(t)\ \textrm{d}t \nonumber\\
&= \delta_e \Delta t\ \sigma^\dagger \sigma + \Omega_k \sigma_x +  \begin{cases}
 \sigma^\dagger\big(\sqrt{\gamma_+ \Delta t}\ e^{\textrm{i}\omega_0 t_d} A_+[k] +  \sqrt{\gamma_-\Delta t}\ e^{-\textrm{i}\omega_0 t_d} A_-[k]\big) + \text{h.c.} & \text{if } k \leq n_d\\
 \sigma^\dagger\big(\sqrt{\gamma_+ \Delta t}\ e^{\textrm{i}\omega_0 t_d}\big(  A_+[k]\cos \theta + e^{\textrm{i}\varphi} A_-[k -2n_d] \sin \theta\big) \\ \ \ \ \ \ \ \ \ \ \ \ \  \ \ \ \ \ \ \  \ \ \ \ \ \ \ \ \ \ \ \ \ + \sqrt{\gamma_- \Delta t} e^{-\textrm{i}\omega_0 t_d}A_-[k]\big) + \text{h.c.} & \text{if } k > n_d
\end{cases},
\end{align}
where $n_d = \left \lfloor t_d / \Delta t \right \rfloor $ and $\Omega_k = \Omega(k\Delta t) \Delta t $. The application of $U[k + 1, k]$ on the matrix product state at time step $k$ requires the implementation of a long-range gate, since it acts on the site corresponding to the two-level system, the $k^\text{th}$ waveguide bin and the $(k - 2n_d)^\text{th}$ waveguide bin. Following the approach introduced in Ref.~\cite{pichler2016photonic}, we implement this long range gate using a sequence of swap operations \cite{shi2006classical} followed by a short-range gate corresponding to $U[k + 1, k]$ \cite{vidal2003efficient}. The update is implemented using the tensor network state python library \texttt{tncontract} \cite{tncontract} along with \texttt{qutip} \cite{johansson2013qutip}\\ \ \\
\noindent We first validate our MPS update implementation \cite{git_code} against the implementation introduced in Ref.~\cite{pichler2016photonic} for an ideal mirror (i.e. $\theta = \pi / 2$). Fig.~\ref{fig:validation_fig} shows a comparison between the two implementation for two distinct settings --- Fig.~\ref{fig:validation_fig}(a) in which the emitter is initially prepared in its excited state and allowed to decay into the waveguide without any external driving, and Fig.~\ref{fig:validation_fig}(b) in which the emitter is initially in its ground state and then driven by an exponentially decaying pulse ($\Omega(t) = \Omega_0 e^{-\alpha t}$). We simulate both of these settings for different mirror phase $\varphi$ --- as is known in such feedback systems with ideal mirrors, a properly chosen mirror phase can result in the emitter not decaying completely into the waveguide mode, rather exciting the \emph{bound state} that exists between the emitter and the waveguide mode. We observe such incomplete decay for $\varphi = 0$, and a complete decay of the emitter into the waveguide mode for other mirror phases. Moreover, the MPS update implementation presented in this section agrees perfectly with the MPS update implementation introduced in Ref.~\cite{pichler2016photonic}. { This perfect agreement can be analytically explained by considering the normal modes of the point-coupling Hamiltonian describing the mirror (as described in section \ref{sec:normal_modes}). Using Eq.~\ref{eq:diag_result} for a perfect mirror ($\theta = \pi / 2$), the two normal modes $b_{L, R}(\omega)$ can be expressed in terms of $a_\pm(x)$ via:
\begin{subequations}
\begin{align}
&b_L(\omega) = \int_{-\infty}^0  a_+(x_+)e^{-\textrm{i}\omega x_+} \frac{\textrm{d}x_+}{\sqrt{2\pi}} - e^{\textrm{i}\varphi} \int_0^\infty a_-(x_-) e^{-\textrm{i}\omega x_-} \frac{\textrm{d}x_-}{\sqrt{2\pi}}, \\
&b_R(\omega) = \int_{-\infty}^0 a_-(x_-) e^{-\textrm{i}\omega x_-} \frac{\textrm{d}x_-}{\sqrt{2\pi}} + e^{-\textrm{i}\varphi} \int_0^\infty a_+(x_+) e^{-\textrm{i}\omega x_+} \frac{\textrm{d}x_+}{\sqrt{2\pi}}.
\end{align}
\end{subequations}
Clearly, $b_L^\dagger(\omega)$ only creates excitations to the left of the mirror (which do not interact with the two-level system) while $b_R^\dagger(\omega)$ only creates excitations to the right of the mirror (which interact with the two-level system). Indeed, using Eq.~\ref{eq:diag_inv_result}, it can easily be seen that the interaction Hamiltonian between the waveguide and the two-level system ($H_\text{wg-TLS}$ defined in Eq.~\ref{eq:wg_tls_hamil}) can be entirely expressed in terms of $b_R(\omega)$:
\begin{align}
H_\text{wg-TLS} = \int_{-\infty}^\infty \bigg[\big(\sqrt{\gamma_+}\sigma^\dagger e^{\textrm{i}(\omega t_d + \varphi)} + \sqrt{\gamma_-}\sigma^\dagger e^{-\textrm{i}\omega t_d}\big)b_R(\omega) + \text{h.c.}\bigg] \frac{\textrm{d}\omega}{\sqrt{2\pi}},
\end{align}
which is identical to the interaction Hamiltonian used in Ref.~\cite{pichler2016photonic}}.\\ \ \\
\begin{figure}[t]
\centering
\includegraphics[scale=0.365]{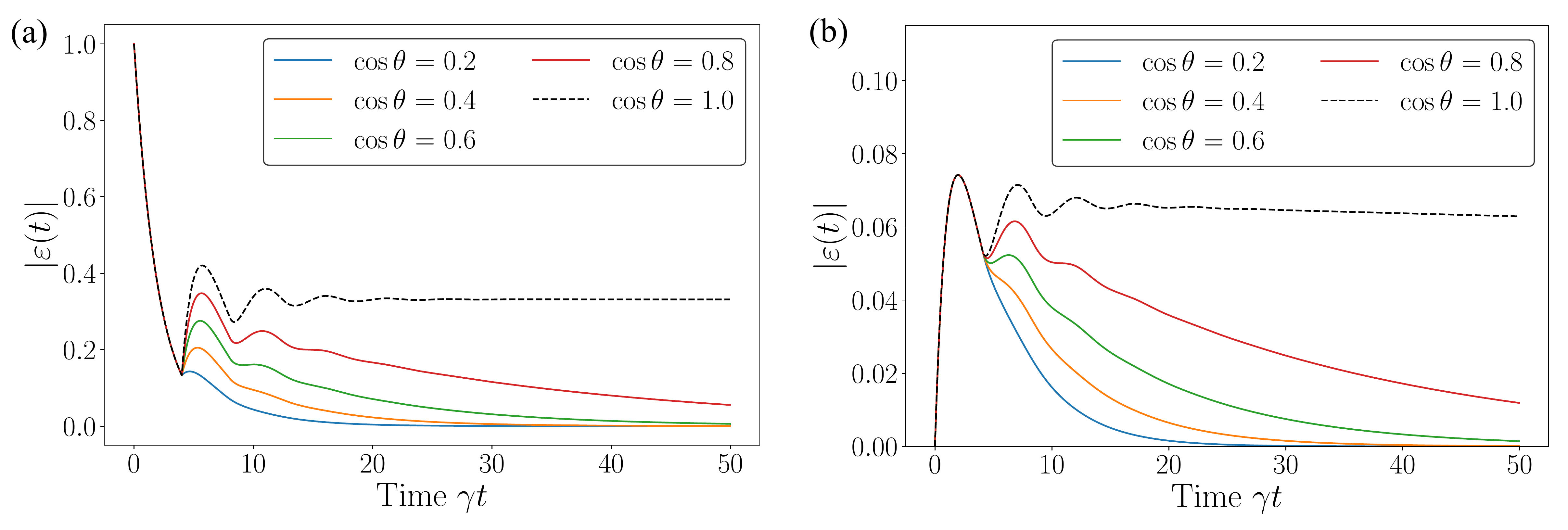}
\caption{Impact of mirror reflectivity on the dynamics of the time-delayed feedback system for two simulation settings: (a) Simulation of an undriven two-level system ($\Omega(t) = 0$) which is initially in its excited state for different mirror reflectivities $\cos \theta$ and (b) simulation of a two-level system initially in its ground state and driven by an exponentially decaying pulse ($\Omega(t) = \Omega_0 e^{-\alpha t}$ for $t > 0$) for different mirror reflectivities $\cos \theta$. $|\varepsilon(t)|$ is the probability amplitude of the two-level system being in the excited state computed using $|\varepsilon(t)|^2 = \langle \sigma^\dagger \sigma \rangle$. It is assumed that $\gamma_+ = \gamma_- = \gamma / 2$, $\delta_e = \omega_e - \omega_0 = 0$, $\omega_0 t_d = \pi$, $\gamma t_d = 2$, $\varphi = 0$ and $\alpha = 2\gamma$. For the discretization into an MPS, we use $\gamma\Delta t = 0.05$, and truncate the dimensionality of the Hilbert space of each waveguide bin to 2 for both forward and backward propagating modes. A threshold of $0.01$ is used in all the Schmidt decompositions performed while applying the swap gates and the short-range gates. { Refer to appendix \ref{app:mps_conv} for convergence studies of the MPS simulations.}}
\label{fig:impact_fig}
\end{figure}
\noindent Next, we study the impact of the non-ideality (i.e.~the mirror reflection being less than 1) in the mirror on the dynamics of the feedback system. Changing the mirror reflection is equivalent to changing the parameter $\theta$ describing the mirror. Fig.~\ref{fig:impact_fig} shows the impact of $\theta$ on the dynamics of the feedback system. Again, we simulate two different settings --- Fig.~\ref{fig:impact_fig}(a) in which the emitter is initially prepared in its excited state and allowed to decay into the waveguide without any external driving, and Fig.~\ref{fig:impact_fig}(b) in which the emitter is initially in its ground state and then driven by an exponentially decaying pulse ($\Omega(t) = \Omega_0 e^{-\alpha t}$). We note that unlike the case of an ideal mirror, having a less than unity reflectivity implies that there is no bound state that the emitter can decay into. Consequently, the emitter always decays to the ground state --- however, as is seen in Fig.~\ref{fig:impact_fig}, the decay rate can be controlled by controlling the reflectivity of the mirror. While this seems reminiscent of the Purcell effect, we note that even with a less than unity reflectivity, the emitter does not decay exponentially into its ground state --- this is a consequence of the non-Markovian nature of the feedback system.\\ \ \\
\noindent { Finally, we point out that an alternative to using the point-coupling Hamiltonian for analyzing the feedback system considered in this section is to directly formulate the Hamiltonian using the normal modes of the mirror-waveguide system. Such a formulation would also be able to capture the impact of the mirror reflectivity on the dynamics of the two-level system if it was accounted for in the construction of the normal modes. The point-coupling Hamiltonian provides an alternative, and equivalent (as shown in section \ref{sec:normal_modes}), approach for modelling linear-optical devices. Apart from the didactic importance of this result, we expect it to be of utility in understanding complicated feedback systems with multiple linear optical devices providing feedback. }

\section{Conclulsion}
This paper resolves the problem of calculating the Hamiltonian for a frequency-independent linear-optical device from its classical scattering matrix. It is shown that an application of the quantum scattering matrix corresponding to the proposed Hamiltonian on an input state is equivalent to applying the inverse of the classical scattering matrix of the linear-optical device on the annihilation operators in the input state. We also diagonalize the point-coupling Hamiltonian and provide a connection between the point-coupling Hamiltonian and the quantization of normal modes of the linear-optical device \cite{dalton1999quasi}. Finally, we demonstrate the practical utility of the proposed Hamiltonian by using it to rigorously formulate an MPS-based update for a time-delayed feedback system wherein the linear-optical device providing feedback is described by a full scattering matrix as opposed to a hard boundary condition.

\section{Acknowledgements}
\noindent R.T.~acknowledges funding from Kailath Stanford Graduate Fellowship.

\bibliography{references.bib}
\newpage
\appendix
{ \section{Solving Heisenberg equations of motion for point coupling Hamiltonian}\label{app:heisenberg}
 \noindent The Heisenberg equation of motion (Eq.~\ref{eq:heisenberg_eq}) can be easily integrated from $t_0$ to $t_0 + \tau$ (where $\tau > 0$) to obtain:
\begin{align}
    a_n(\omega; t_0 + \tau) = a_n(\omega; t_0)\exp(-\textrm{i}\omega \tau)-\textrm{i}\sum_{m=1}^N \text{V}_{n, m}e^{-\textrm{i}\omega x_n^0} \int_{t_0}^{t_0 + \tau} a_m(x_m^0; t')e^{-\textrm{i}\omega (t_0 + \tau - t')} \frac{\textrm{d}t' }{\sqrt{2\pi}}
\end{align}
Therefore, it follows from Eq.~\ref{eq:pos_annihilation_op} that $a_n(x_n; t_0 + \tau)$ is given by:
\begin{align}\label{eq:integrated_heisenberg}
    a_n(x_n; t_0 + \tau) &= a_n(x_n - \tau; t_0)-\textrm{i}\sum_{m=1}^N \text{V}_{n, m}\int_{t_0}^{t_0 + \tau} a_m(x_m^0; t') \delta(x_n - x_n^0 - t_0 - \tau + t')\textrm{d}t' \nonumber \\
              &= a_n(x_n - \tau; t_0) - \textrm{i}\bigg[\sum_{m=1}^N \text{V}_{n, m}a_m(x_m^0; t_0 + \tau- (x_n - x_n^0))\bigg]\Theta(x_n^0 \leq x_n \leq x_n^0 + \tau)
\end{align}
where  the function $\Theta(x_1 \leq x \leq x_2)$ is defined in Eq.~\ref{eq:step_fun}. Imposing Eq.~\ref{eq:integrated_heisenberg} at $x_n = x_n^0 \ \forall \ n \in \{1, 2 \dots N\}$, we obtain:
\begin{align}
    a_n(x_n^0; t_0 + \tau) = a_n(x_n^0 - \tau; t_0) -\frac{\textrm{i}}{2}\sum_{m=1}^N \text{V}_{n, m}a_m(x_m^0; t_0 + \tau)  \ \forall \ n \in \{1, 2 \dots N\}
\end{align}
from which we can obtain $a_n(x_n^0; t)$ in terms of operators at $t = t_0$:
\begin{align}\label{eq:boundary_cond}
    \begin{bmatrix}
    a_1(x_1^0; t_0 + \tau) \\
    a_2(x_2^0; t_0 + \tau) \\
    \vdots \\
    a_N(x_N^0; t_0 + \tau)
    \end{bmatrix} = \bigg(\textbf{I} + \frac{\textrm{i}\textbf{V}}{2} \bigg)^{-1}\begin{bmatrix}
    a_1(x_1^0 - \tau; t_0) \\
    a_2(x_2^0 - \tau; t_0) \\
    \vdots \\
    a_N(x_N^0 - \tau; t_0)
    \end{bmatrix}
\end{align}
where $\textbf{V}$ is a $N \times N$ Hermitian matrix formed by $\text{V}_{m, n}$ as its elements and $\textbf{I}$ is the identity matrix of size $N$. Substituting Eq.~\ref{eq:boundary_cond} into Eq.~\ref{eq:integrated_heisenberg} together with the substitution $x_n = x_n^0 + y$, we obtain Eq.~\ref{eq:sol_heisenberg}

\section{Calculating normal modes of the point coupling Hamiltonian}\label{app:diagonalization_details}
\noindent Since the normal mode annihilation operator $b_n(\omega)$ is expressible as a linear combination of the annihilation operators $a_n(x_n)$, we assume the following ansatz for $b_n(\omega)$:
\begin{align}\label{eq:ansatz}
b_n(\omega) = \sum_{m=1}^N \int_{-\infty}^\infty \mathcal{F}_{n, m}(\omega, x) a_m(x + x_m^0)\frac{ \textrm{d}x}{\sqrt{2\pi}},
\end{align}
where $\mathcal{F}_{n, m}(\omega, x)$ is a function that is to be determined. Using Eq.~\ref{eq:comm_with_hamil}, we obtain the following differential equation for $\mathcal{F}_{n, m}(\omega, x)$:
\begin{align}\label{eq:normal_modes_pde}
\textrm{i}\frac{\partial \boldsymbol{\mathcal{F}}(\omega, x)}{\partial x} + \delta(x) \boldsymbol{\mathcal{F}}(\omega, 0) \textbf{V} = \omega \boldsymbol{\mathcal{F}}(\omega, x).
\end{align}
Here $\boldsymbol{\mathcal{F}}(\omega, x)$ is a $N \times N$ complex matrix with the functions $\mathcal{F}_{n, m}(\omega, x)$ as its elements. The solution for Eq.~\ref{eq:normal_modes_pde} is of the form:
\begin{align}
\boldsymbol{\mathcal{F}}(\omega, x) = \begin{cases}
\boldsymbol{\mathcal{F}}_+(\omega) e^{-\textrm{i}\omega x} & x > 0 \\
\boldsymbol{\mathcal{F}}_-(\omega) e^{-\textrm{i}\omega x} & x < 0
\end{cases}.
\end{align}
To evaluate the matrices $\mathcal{F}_+(\omega)$ and $\mathcal{F}_-(\omega)$, we use the boundary condition obtained on integrating Eq.~\ref{eq:normal_modes_pde} across a small interval around $x = 0$:
\begin{align}
-\textrm{i}\big[\boldsymbol{\mathcal{F}}(\omega, x = 0^+) -\boldsymbol{\mathcal{F}}(\omega, x = 0^-)  \big] + \frac{1}{2}\big[\boldsymbol{\mathcal{F}}(\omega, x = 0^+) +\boldsymbol{\mathcal{F}}(\omega, x = 0^-)  \big] \textbf{V}= 0,
\end{align}
from which it immediately follows that $\boldsymbol{\mathcal{F}}_+(\omega) = \boldsymbol{\mathcal{F}}_-(\omega)\textbf{S}^\dagger$, where $\textbf{S}$ is the classical scattering matrix of the linear-optical device. Moreover, using Eq.~\ref{eq:normal_modes_pde} along with Eq.~\ref{eq:comm_with_itself}, we obtain $\boldsymbol{\mathcal{F}}_-^\dagger(\omega) \boldsymbol{\mathcal{F}}_-(\omega) = \textbf{I}$ -- any choice for $\boldsymbol{\mathcal{F}}(\omega)$ is valid as long as it satisfies this constraint. Choosing $\boldsymbol{\mathcal{F}}_-(\omega) = \textbf{I}$ and $\boldsymbol{\mathcal{F}}_+(\omega) = \textbf{S}^\dagger$, we obtain Eq.~\ref{eq:diag_result}.\\ \ \\
\noindent To derive Eq.~\ref{eq:diag_inv_result}, we assume the following anstaz for $a_n(x_n)$ in terms of $b_n(\omega)$:
\begin{align}\label{eq:ansatz_inv}
a_n(x + x_n^0) = \int_{-\infty}^\infty\sum_{n, m} \mathcal{G}_{n, m}(x, \omega) b_m(\omega) \frac{\textrm{d}\omega}{\sqrt{2\pi}},
\end{align}
where $\mathcal{G}_{n, m}(x, \omega)$ is a function that is to be determined. Note that $\mathcal{G}_{n, m}(x, \omega) = \sqrt{2\pi} [a_n(x + x_n^0), b_m^\dagger(\omega)]$ and therefore from Eq.~\ref{eq:ansatz} it follows that $\mathcal{G}_{n, m}(x, \omega) = \mathcal{F}_{m, n}^*(\omega, x)$. We thus immediately obtain:
\begin{align}
\mathcal{G}_{n, m}(x, \omega) =
\begin{cases}
\delta_{n, m} e^{\textrm{i}\omega(x - x_n^0)} & \text{if } x < x_n^0 \\
\text{S}_{n, m} e^{\textrm{i}\omega (x - x_n^0)} & \text{if } x > x_n^0
\end{cases}.
\end{align}
Substituting this into Eq.~\ref{eq:ansatz_inv}, we immediately obtain Eq.~\ref{eq:diag_inv_result}.

\section{Convergence of Matrix-product-state simulations}\label{app:mps_conv}
\noindent Simulating time-delayed feedback with MPS, as described in section \ref{sec:mps}, requires discretizing the waveguide modes which are described as continuas of harmonic oscillators into waveguide bins (Eq.~\ref{eq:disc}). This introduces a simulation parameter $\delta t$ which controls the coarseness of this discretization. Moreover, application of a unitary impacting multiple bins in the MPS requires us to perform Schmidt decompositions on the state, and neglect Schmidt vectors with coefficients smaller than a specified tolerance. For the MPS-based simulation to be accurate, it is necessary to ensure that it has converged with respect to these two parameters. In this appendix, we present convergence studies with respect to these parameters to show that the choice of $\Delta t$ ($=0.05/\gamma$) and tolerance (0.01) assumed in section \ref{sec:mps} indeed results in accurate simulations.\\ \ \\
\noindent For the convergence study, we study the problem of an emitter decaying into the waveguide with a time-delayed feedback provided by the mirror and calculate the probability amplitude $|\varepsilon(t)|$ of the two-level system being in its excited state at time $t$ calculated using $|\varepsilon(t)|^2 = \langle \sigma^\dagger \sigma \rangle$. We also assume that the mirror is perfect with phase $\varphi = 0$ (which, as can be seen from Fig.~\ref{fig:validation_fig}(a) results in the two-level system not completely decaying into the ground state). In this case, the time-dependence of the complex amplitude of the excited state $\varepsilon(t)$ is given by the following ordinary differential equation (ODE) \cite{tufarelli2013dynamics}:
\begin{align}\label{eq:ode}
\frac{\textrm{d}\varepsilon(t)}{\textrm{d}t} = - \frac{\gamma}{2}\varepsilon(t) - \frac{\gamma}{2} e^{\textrm{i}(2\omega_0 t_d + \varphi)}\varepsilon(t - 2t_d),
\end{align}
where $\gamma = \gamma_+ + \gamma_-$ is the total decay rate of the two-level system into the forward and backward propagating waveguide modes and this equation is solved subject to the initial condition $\varepsilon(0) = 1$. $\varepsilon(t)$ obtained on solving this ODE thus provides a benchmark simulation that the MPS simulation can be compared against to gauge its convergence.\\ \ \\
Fig.~\ref{fig:conv_fig} shows the results of the convergence studies. The dependence of $|\varepsilon(t)|$ obtained from the MPS update on the discretization time-step $\Delta t$ is shown in Fig.~\ref{fig:conv_fig}(a) --- it can be seen that choosing $\Delta t< 0.15 / \gamma$ is sufficient to ensure that the MPS simulation has converged and agrees with the ODE simulation. A similar study performed with respect to the tolerance used in the Schmidt decomposition is shown in Fig.~\ref{fig:conv_fig}(b) from which it can be seen that a tolerance smaller than 0.1 is sufficient to ensure that the MPS simulation has converged and agrees with the ODE simulation. 

\begin{figure}
\centering
\includegraphics[scale=0.37]{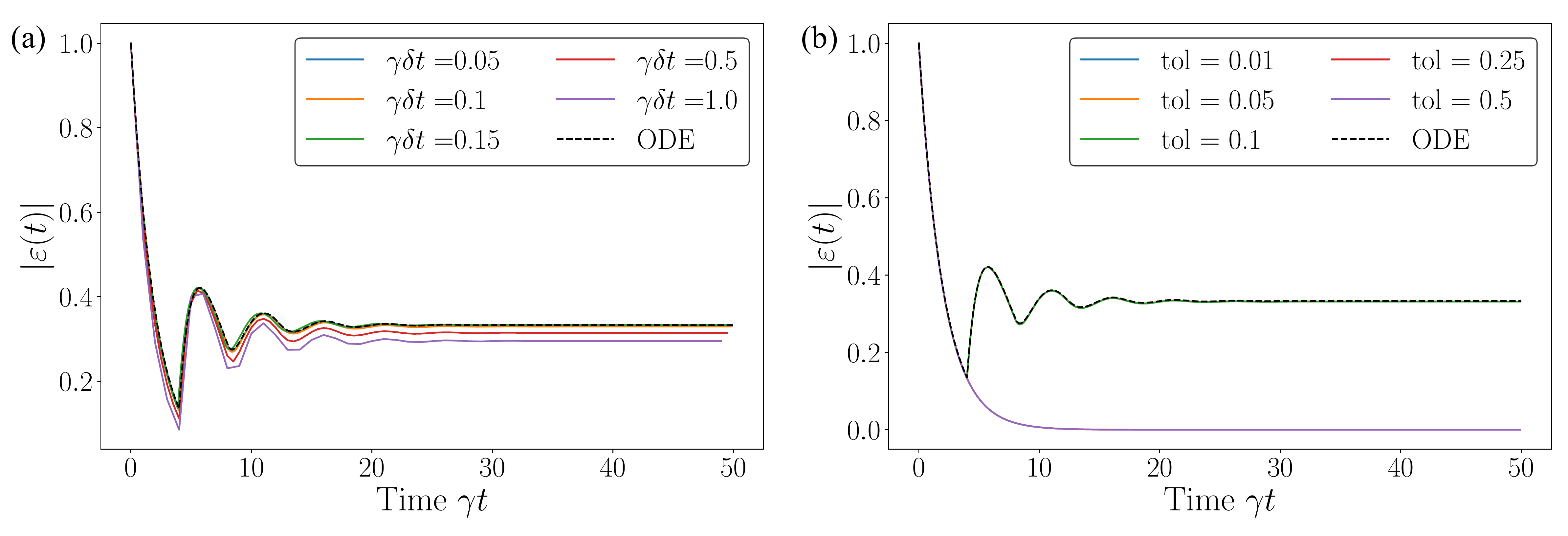}
\caption{{ Convergence studies on the MPS simulations for the time-delayed feedback system introduced in section \ref{sec:mps}. The undriven two-level system ($\Omega(t) = 0$) is initially in its excited state and is allowed to decay into the forward and backward propagating waveguide modes with an ideal mirror ($\theta = \pi / 2, \varphi=0$) providing feedback. $|\varepsilon(t)$ simulated using MPS update for (a) different discretization time-steps $\Delta t$ and (b) different tolerances that govern the number of Schmidt vectors retained after every Schmidt decomposition on the MPS. The dotted black line shows $|\varepsilon(t)|$ obtained on solving the ODE (Eq.~\ref{eq:ode}) with a very small time-step ($\gamma \Delta t = 0.001$). $|\varepsilon(t)|$ is the probability amplitude of the two-level system being in the excited state computed using $|\varepsilon(t)|^2 = \langle \sigma^\dagger \sigma \rangle$. In the simulations shown in (a) the Schmidt tolerance is assumed to be 0.01 and in the simulations shown in (b) $\gamma\Delta t = 0.05$. It is also assumed that $\gamma_+ = \gamma_- = \gamma / 2$, $\delta_e = \omega_e - \omega_0 = 0$, $\omega_0 t_d = \pi$ and $\gamma t_d = 2$. }}
\label{fig:conv_fig}
\end{figure}

}

\end{document}